\documentclass{article}
\usepackage{spconf,amsmath,amssymb,graphicx,hyperref,booktabs}
\usepackage[utf8]{inputenc}
\usepackage{CJKutf8}

\newcommand{\kanacer}{Kana-CER}
\newcommand{\jp}[1]{\begin{CJK}{UTF8}{min}#1\end{CJK}}

\title{Pronunciation-Oriented Reinforcement Learning for Japanese Text-to-Speech with Kana-Domain ASR Rewards}

\name{Shiao Zhu$^{*}$, Lianbo Liu$^{*}$, Kai Washizaki, Koki Nikaido, Yui Sudo}

\address{SB Intuitions Corp., Tokyo, Japan\\
$^{*}$Equal contribution.}

\begin{document}
\maketitle


\begin{abstract}
Character error rate (CER) computed by automatic speech recognition (ASR) is widely used as an intelligibility reward for reinforcement learning (RL) post-training of text-to-speech (TTS) systems. For Japanese, however, orthographic CER introduces a representation mismatch for pronunciation-oriented optimization: distinct kanji readings may collapse to the same orthographic representation, while equivalent pronunciations may admit different orthographic forms. We instead compute CER in the kana domain using a kana-transcribing ASR model and reference readings (\kanacer{}). Under matched group relative policy optimization (GRPO) conditions, \kanacer{} reduces target-kanji reading error by approximately $26\%$ relative to the orthographic CER reward, while maintaining comparable orthographic CER and similar speaker similarity and objective speech quality. It also reaches its best validation performance in substantially fewer optimization steps ($4$k vs.\ $18$k). We further observe severe output elongation under unregularized \kanacer{} optimization, which is substantially suppressed by KL regularization.
\end{abstract}

\begin{keywords}
text-to-speech, reinforcement learning, reward design, Japanese
pronunciation, GRPO
\end{keywords}

\section{Introduction}
\label{sec:intro}

Recent work has explored reinforcement learning (RL) for text-to-speech (TTS) post-training, where generated speech is transcribed by automatic speech recognition (ASR) and the character error rate (CER) or word error rate (WER) between the resulting transcript and the conditioning text is used as an intelligibility reward \cite{sun2025f5rtts,liu2025grpotts}.

For Japanese, however, orthography and pronunciation are not in one-to-one correspondence, making orthographic CER only an indirect proxy for pronunciation accuracy. Kanji, the logographic characters used in Japanese writing, often have multiple readings whose selection depends on lexical context. For example, \jp{固} is read \jp{かた}~(\textit{kata}) in \jp{固める} but \jp{コ}~(\textit{ko}) in \jp{固定}. Correct reading selection remains a long-standing problem in Japanese TTS \cite{kakegawa2021phonetic,hida2022polyphone}. When CER is computed after orthographic ASR transcription, such pronunciation distinctions may be obscured: an ASR system may recover the intended word despite a mispronunciation, masking the error in the transcript \cite{park2026intent}. Conversely, the same pronunciation may admit different orthographic forms, so an ASR substitution between such forms can increase CER even when the pronunciation itself is correct \cite{liu2026sarashina22tts}. Orthographic CER may therefore both miss pronunciation errors and penalize orthographic differences that are irrelevant to pronunciation.

Pronunciation-sensitive rewards have recently been explored for Chinese and English TTS post-training, including phone-level rewards in SwanVoice \cite{swanvoice} and phone--tone accuracy rewards in SwanTale \cite{swantale}, which builds on group relative policy optimization (GRPO) \cite{shao2024deepseekmath}. In Japanese TTS, context-dependent kanji readings pose a particular challenge for rewards based on orthographic ASR, which may not faithfully reflect pronunciation accuracy. Kana makes these reading distinctions explicit and therefore provides a natural representation for pronunciation-oriented optimization. Building on \kanacer{}, originally introduced as an evaluation
metric for Japanese kanji-reading accuracy
\cite{liu2026sarashina22tts}, we develop a pronunciation-oriented
RL post-training approach for Japanese TTS that uses kana-domain
ASR feedback to target context-dependent kanji-reading errors. To our knowledge, this is the first study to investigate kana-domain ASR rewards for pronunciation-oriented RL post-training of Japanese TTS.

Our contributions are as follows:
\begin{itemize}
\item We formulate a kana-domain ASR reward for Japanese TTS post-training that makes reading distinctions explicit in the reward computation
(Sec.~\ref{sec:reward}).

\item We evaluate the kana-domain reward on the Joyo Kanji Yomi Benchmark\footnote{
\url{https://github.com/sbintuitions/Joyo-Kanji-Yomi-Benchmark}
}, where it achieves a $26\%$ relative reduction in target-kanji reading error compared with the orthographic CER reward while maintaining comparable orthographic CER, and reaches its best validation performance in substantially fewer optimization steps ($4$k vs.~$18$k)
(Secs.~\ref{sec:main}--\ref{sec:dynamics}).

\item We identify severe output elongation under unregularized \kanacer{} optimization and show that KL regularization substantially suppresses this failure mode, while speaker similarity and objective speech quality remain similar across the evaluated systems
(Secs.~\ref{sec:hack}--\ref{sec:quality}).
\end{itemize}

\section{Method}
\subsection{TTS generation and ASR-derived rewards}
\label{sec:reward}

We use the language model of Sarashina2.2-TTS
\cite{liu2026sarashina22tts} as the TTS policy $\pi_\theta$.
Given conditioning text $x$ and a reference audio prompt $p$, the policy
autoregressively generates a sequence of semantic speech tokens:
\begin{equation}
  z \sim \pi_\theta(\cdot \mid x,p).
  \label{eq:generation}
\end{equation}
The generated sequence $z$ is converted into a waveform $g$ by the
acoustic decoder and neural vocoder. During RL post-training, only
$\pi_\theta$ is updated; the decoder and vocoder remain frozen.

We derive the reward by transcribing $g$ with ASR and comparing the
transcript with a reference representation of $x$. Throughout,
$\mathrm{CER}(a,b)$ denotes the character-level edit error rate between
reference $a$ and hypothesis $b$, normalized by $|a|$, with punctuation,
symbols, and whitespace removed. The conventional orthographic CER reward measures the error as
\begin{equation}
  e_{\mathrm{orth}}
  =
  \mathrm{CER}\bigl(x,\mathcal{A}_{\mathrm{orth}}(g)\bigr),
  \label{eq:orth_error}
\end{equation}
where $\mathcal{A}_{\mathrm{orth}}$ is an orthographic Japanese ASR model.
Our kana-domain reward instead measures the error in the pronunciation
domain:
\begin{equation}
  e_{\mathrm{kana}}
  =
  \mathrm{CER}\bigl(k(x),\mathcal{A}_{\mathrm{kana}}(g)\bigr),
  \label{eq:kana_error}
\end{equation}
where $k(x)$ is the reference kana reading and
$\mathcal{A}_{\mathrm{kana}}$ is a Whisper-based ASR model
\cite{liu2026sarashina22tts} fine-tuned to output katakana.

Either error is mapped to a bounded reward:
\begin{equation}
  r_d = 1-\tanh(\alpha\,e_d),
  \qquad d\in\{\mathrm{orth},\mathrm{kana}\},
  \label{eq:reward}
\end{equation}
where $\alpha>0$ controls how rapidly the reward decreases with the error.
\subsection{GRPO post-training}
\label{sec:grpo}

For each input, we sample $G$ completions from the rollout policy and
score the resulting speech using one of the two rewards defined in
Sec.~\ref{sec:reward}. The rewards are standardized within each group
to obtain group-relative advantages:
\begin{equation}
A_i =
\frac{r_i-\bar{r}}{\sigma_r+\epsilon},
\label{eq:advantage}
\end{equation}
where $r_i$ is the reward of the $i$-th completion, $\bar{r}$ and $\sigma_r$ are the mean and standard deviation of
the rewards within the group, and $\epsilon$ is a small constant for
numerical stability.

We optimize the policy with the objective
\begin{equation}
  \mathcal{L}(\theta)
  =
  \mathcal{L}_{\mathrm{policy}}(\theta)
  +
  \beta\,\widehat{D}_{\mathrm{KL}}
  \bigl(\pi_\theta \,\|\, \pi_{\mathrm{ref}}\bigr),
  \label{eq:grpo_objective}
\end{equation}
where $\mathcal{L}_{\mathrm{policy}}$ is the clipped GRPO policy loss
computed from these advantages, $\pi_{\mathrm{ref}}$ is the frozen
pre-RL reference policy, and $\beta$ controls the strength of KL
regularization. Both reward conditions use the same policy optimization,
sampling procedure, and hyperparameters; only the ASR-derived reward
signal differs.

\begin{table*}[t]
\centering
\caption{Comparison of pre-RL, SFT, and GRPO models. Values are mean $\pm$ std across five generation seeds; all error rates are in \% and lower is better. \emph{cap@1} denotes \kanacer{}~(kanji) with per-sample CER clamped to 1.0 before averaging.}
\label{tab:main}

\setlength{\tabcolsep}{3.5pt}

\begin{tabular}{@{}lc|ccc|c|ccc@{}}
\toprule
& &
\multicolumn{3}{c}{Kana-Whisper} &
\multicolumn{1}{c}{Whisper} &
\multicolumn{3}{c}{wav2vec2-hiragana} \\
\cmidrule(lr){3-5}
\cmidrule(lr){6-6}
\cmidrule(lr){7-9}

Method & RL Step
& \kanacer{} & \kanacer{} & \kanacer{} & CER
& \kanacer{} & \kanacer{} & \kanacer{} \\

&
& (kanji)$\downarrow$ & (cap@1)$\downarrow$ & (sent.)$\downarrow$ & $\downarrow$
& (kanji)$\downarrow$ & (cap@1)$\downarrow$ & (sent.)$\downarrow$ \\

\midrule

pre-RL & ---
& 10.77{\small$\pm$0.57}
& 6.97{\small$\pm$0.08}
& 4.59{\small$\pm$0.85}
& 6.08{\small$\pm$0.58}
& 16.69{\small$\pm$0.36}
& 10.90{\small$\pm$0.21}
& 8.62{\small$\pm$0.18} \\
\midrule
SFT & ---
& 7.61{\small$\pm$0.13}
& 5.48{\small$\pm$0.08}
& 3.38{\small$\pm$0.83}
& 5.20{\small$\pm$0.38}
& 13.77{\small$\pm$0.15}
& 9.52{\small$\pm$0.11}
& 8.04{\small$\pm$0.15} \\

RL w/ CER & 18000
& 9.65{\small$\pm$0.12}
& 6.51{\small$\pm$0.09}
& 2.44{\small$\pm$0.06}
& \textbf{4.10}{\small$\pm$0.16}
& 14.78{\small$\pm$0.33}
& 9.71{\small$\pm$0.18}
& 7.01{\small$\pm$0.11} \\

RL w/ \kanacer{} & 4000
& \textbf{7.17}{\small$\pm$0.07}
& \textbf{5.16}{\small$\pm$0.11}
& \textbf{2.19}{\small$\pm$0.03}
& 4.15{\small$\pm$0.04}
& \textbf{11.52}{\small$\pm$0.28}
& \textbf{7.94}{\small$\pm$0.10}
& \textbf{6.54}{\small$\pm$0.10} \\

\bottomrule
\end{tabular}
\end{table*}
\section{Experimental setup}
\label{sec:setup}

\paragraph*{Data and model.}
We initialize GRPO from the Sarashina2.2-TTS Stage~1 checkpoint, which is obtained after large-scale TTS pre-training and serves as the common initialization for supervised fine-tuning (SFT)  and RL. Stage~2 applies pronunciation-oriented SFT to Stage~1 using targeted synthetic data \cite{liu2026sarashina22tts}. These data were generated for specific kanji–reading pairs from the Joyo Kanji List (Japan’s official list of common-use kanji), with each sentence providing a context that determines the intended reading.
For RL post-training, we use a subset of these data containing
$129{,}965$ utterances covering $4{,}362$ kanji--reading pairs and
$2{,}129$ distinct kanji, with up to $30$ sentences per pair.
Each sentence is accompanied by its reference kana pronunciation,
which is used to compute the \kanacer{} reward. Zero-shot audio
prompts are randomly assigned once and shared across reward conditions. For reward computation, we use Whisper \cite{radford23a}
large-v3-turbo%
\footnote{\url{https://huggingface.co/openai/whisper-large-v3-turbo}}
as $\mathcal{A}_{\mathrm{orth}}$ and Kana-Whisper
\cite{liu2026sarashina22tts} as $\mathcal{A}_{\mathrm{kana}}$.
Kana-Whisper achieves a \kanacer{} of $0.979\%$ on real JSUT recordings \cite{takamichi2020jsut,liu2026sarashina22tts}, supporting its use as a pronunciation-sensitive reward model.

\paragraph*{Evaluation.}
We use the Joyo Kanji Yomi Benchmark introduced in \cite{liu2026sarashina22tts}, which systematically covers context-dependent readings of Joyo kanji, including less common readings. Each sentence is annotated with the target reading span in the reference kana sequence. Our primary metric, \textbf{\kanacer{} (kanji)}, computes kana error only over this annotated span, thereby isolating errors in realizing the intended kanji reading from transcription errors elsewhere in the sentence. We additionally report
sentence-level \textbf{\kanacer{} (sent.)} and standard orthographic
\textbf{CER}, with the latter computed using the same Whisper large-v3-turbo model used for the orthographic reward. To test whether the observed improvements are specific to the ASR model used for the kana-domain reward, we additionally evaluate the same outputs using an independent wav2vec2-based
hiragana ASR model\footnote{\url{https://huggingface.co/sakasegawa/japanese-wav2vec2-large-hiragana-ctc}}
that is not used during RL training, with its hiragana transcripts normalized
to katakana before CER computation. For output-duration analysis, \emph{Dur./base} denotes the ratio of
mean synthesized duration to that of the pre-RL model, while
\emph{Paired} denotes the mean utterance-level duration ratio computed
over matched sentences.

We split the benchmark test set into 20\% validation and 80\% reporting subsets. Checkpoints are evaluated every 1k steps and selected by the mean of validation orthographic CER and sentence-level \kanacer{}; final results use the reporting subset. To limit rare ASR-collapse artifacts,
samples with five or more consecutive repetitions of a single character in
the ASR transcript ($<0.1\%$) are excluded from error-rate averaging.
\paragraph*{RL implementation.}
We use $G=16$ completions per input, a per-device train batch of $12$
completions on eight GPUs, a learning rate of $3\times10^{-6}$, temperature $0.9$, top-$p$ $0.95$, and a
maximum completion length of $750$ tokens. Training uses bf16 precision, $\alpha=3$ for the reward transformation, and
$\beta=0.04$ for KL regularization unless otherwise stated. We build on the GRPO implementation in TRL~1.7.0 \cite{vonwerra2020trl} with vLLM-based generation, and use Dr.~GRPO loss aggregation \cite{liu2025understanding} and a token-level importance-sampling correction for the mismatch between the vLLM sampling and training log-probabilities. All main reward-comparison results are averaged
over five generation seeds.
\section{Results}

\begin{table*}[t]
\centering
\caption{Output duration, speaker similarity, and objective speech
quality of the selected checkpoints. Values are five-seed mean $\pm$ std.}
\label{tab:quality}

\setlength{\tabcolsep}{3pt}
\begin{tabular*}{\textwidth}{@{\extracolsep{\fill}}lcc|c|cccc@{}}

\toprule
&
\multicolumn{2}{c}{Output duration} &
\multicolumn{1}{c}{Speaker similarity} &
\multicolumn{4}{c}{Objective speech quality} \\
\cmidrule(lr){2-3}
\cmidrule(lr){4-4}
\cmidrule(lr){5-8}

Method
& Dur./base
& Paired
& SIM-o
& UTMOS-s
& UTMOS-v2
& P.808
& OVRL \\

\midrule

pre-RL
& 1.000
& 1.000
& 74.36{\small$\pm$0.26}
& 3.183{\small$\pm$0.008}
& 2.903{\small$\pm$0.014}
& 3.808{\small$\pm$0.010}
& 3.240{\small$\pm$0.005} \\

RL w/ CER
& 0.971{\small$\pm$0.07}
& 0.986{\small$\pm$0.08}
& 74.29{\small$\pm$0.09}
& 3.229{\small$\pm$0.009}
& 2.933{\small$\pm$0.014}
& 3.836{\small$\pm$0.007}
& 3.252{\small$\pm$0.008} \\

RL w/ \kanacer{}
& 0.985{\small$\pm$0.07}
& 1.000{\small$\pm$0.08}
& 74.24{\small$\pm$0.09}
& 3.252{\small$\pm$0.008}
& 2.913{\small$\pm$0.003}
& 3.833{\small$\pm$0.006}
& 3.252{\small$\pm$0.004} \\

\bottomrule
\end{tabular*}
\end{table*}
\subsection{Kanji-reading accuracy across reward conditions}
\label{sec:main}
\begin{figure}[!ht]
  \centering
  \includegraphics[width=\columnwidth]{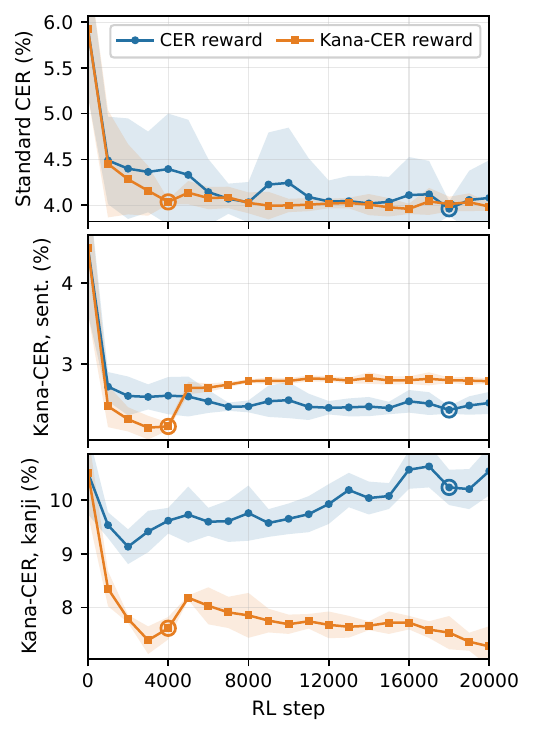}
\caption{Evaluation metrics during RL on the validation split with
  $\beta=0.04$. Curves show the mean across five generation seeds with
$\pm 1\sigma$ shading. Rings indicate the checkpoints selected by the common
  validation criterion. The pre-RL model is shown at step~0.}
  \label{fig:curves}
\end{figure}
The main effect of the \kanacer{} reward is a substantial improvement in the intended kanji readings. With the Kana-Whisper evaluator, \kanacer{} (kanji) decreases from $9.65\%$ under the orthographic CER reward to $7.17\%$, a $25.7\%$ relative reduction. This result indicates that the kana-domain reward better captures pronunciation distinctions that may be lost after orthographic transcription. The independent w2v2-hiragana evaluator shows the same trend, confirming that the gain is not specific to the Kana-Whisper model used in the reward pipeline.

Importantly, orthographic transcription accuracy remains comparable between the two reward conditions, with CER values of $4.10\%$ and $4.15\%$. This is consistent with the two rewards sharing the same underlying objective of accurately realizing the conditioning text, while differing in the representation used to measure that accuracy.

Although our RL training uses only roughly half of the synthetic
pronunciation data used for Stage~2 SFT, the \kanacer{}-optimized
model achieves lower error than Stage~2 SFT on every metric in
Table~\ref{tab:main}.

\subsection{Training dynamics under different reward objectives}
\label{sec:dynamics}

\begin{figure}[t]
\centering
\includegraphics[width=\columnwidth]{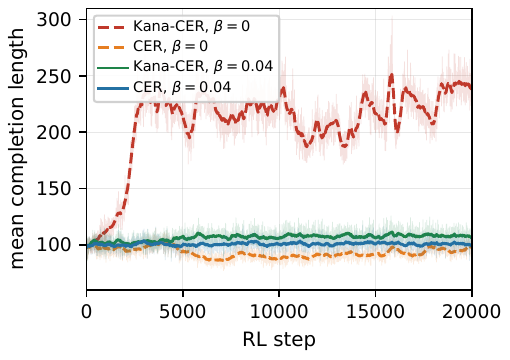}
\caption{Mean completion length during RL under the CER and \kanacer{}
rewards with and without KL regularization. Dashed: $\beta=0$; solid:
$\beta=0.04$. The pale band shows the raw fluctuation across training
rollouts and the bold curve its moving average.}
\label{fig:degen}
\end{figure}
Fig.~\ref{fig:curves} shows distinct optimization dynamics under the two reward objectives. The \kanacer{} reward improves target-kanji reading accuracy substantially earlier than the orthographic CER reward and maintains lower target-reading error over subsequent checkpoints. Under the common validation criterion defined in Sec.~\ref{sec:setup}, \kanacer{} reaches its best validation performance at step $4$k, whereas the orthographic CER reward reaches its optimum at step $18$k. This earlier optimum is consistent with the kana-domain reward retaining pronunciation distinctions that may be discarded by orthographic transcription, resulting in more efficient pronunciation-oriented optimization.
\subsection{Length degeneration and KL regularization}
\label{sec:hack}

The two reward objectives exhibit markedly different length dynamics in the absence of KL regularization. As shown in Fig.~\ref{fig:degen}, with $\beta=0$, the orthographic CER reward causes only a moderate change in completion length, whereas \kanacer{} induces rapid and severe output elongation, with the mean completion length more than doubling within the first few thousand steps.
Under the regularized setting used in the main experiments ($\beta=0.04$), this elongation is substantially suppressed. The selected \kanacer{} checkpoint remains close to the pre-RL model in output duration on the Joyo Kanji Yomi Benchmark as shown in Table~\ref{tab:quality}.

\subsection{Speaker similarity and objective quality}
\label{sec:quality}

To assess whether the pronunciation gains affect other aspects of speech generation, we evaluate the pre-RL model and the two selected
GRPO checkpoints from Table~\ref{tab:main} on the 500-utterance Japanese subset of CV3-Eval \cite{du2025cosyvoice3}. Speaker similarity is measured by SIM-o using ERes2Net \cite{chen2023eres2net}, while objective speech quality is evaluated with UTMOS \cite{saeki2022utmos,baba2024utmosv2} and DNSMOS \cite{reddy2021dnsmos,reddy2022dnsmos835}. As shown in Table~\ref{tab:quality}, the three systems obtain similar scores across all five metrics, with no apparent degradation in speaker similarity or objective speech quality.

\section{Conclusion}

We investigated a kana-domain ASR reward for RL post-training of Japanese TTS, motivated by the mismatch between orthographic representation and pronunciation. On the Joyo Kanji Yomi Benchmark, the \kanacer{} reward reduces target-kanji reading error by approximately $26\%$ compared with the orthographic CER reward under matched GRPO conditions, while maintaining comparable orthographic CER and similar speaker similarity and objective speech quality. It also reaches its best validation performance substantially earlier ($4$k vs.~$18$k steps).
We further observe severe output elongation under unregularized \kanacer{} optimization, a failure mode substantially suppressed by KL regularization. The current study uses one $0.5$B base model and requires reference pronunciations during training; broader validation across model scales and TTS architectures remains future work.
\vfill\pagebreak

\bibliographystyle{IEEEbib}
\bibliography{refs}

\end{document}